\documentclass[journal=ancac3,manuscript=article]{achemso}

\author{Zi-Yu Chen}
\altaffiliation{these authors contributed equally to this work.}
\email{ziyuch@caep.ac.cn}

\author{Rui Qin}
\altaffiliation{these authors contributed equally to this work.}
\email{qinrui.phy@outlook.com}

\affiliation[caep]
{National Key Laboratory of Shock Wave and Detonation Physics, Institute of Fluid Physics, China Academy of Engineering Physics, Mianyang 621999, China}

\title{Strong-field nonlinear optical properties of monolayer black phosphorus}

\keywords{Black phosphorus; 2D materials; strong field; high harmonic generation;}

\begin{document}

\begin{abstract}
Within the past few years, atomically thin black phosphorus (BP) has been demonstrated as a fascinating new 2D material that is promising for novel nanoelectronics and nanophotonics applications, due to its many unique properties such as direct and widely tunable bandgap, high carrier mobility and remarkable intrinsic in-plane anisotropy. However, its important extreme nonlinear behavior and ultrafast dynamics of carriers under strong-field excitation have yet to be revealed to date. Herein, we report nonperturbative high harmonic generation (HHG) in monolayer BP by first-principles simulations. We show that BP exhibits extraordinary HHG properties, with clear advantages over three major types of 2D materials under intensive study, i.e., semimetallic graphene, semiconducting MoS$_2$, and insulating hexagonal boron nitride, in terms of HHG cutoff energy and spectral intensity. This study advances the scope of current research activities of BP into a new regime, suggesting its promising future in applications of extreme-ultraviolet and attosecond nanophotonics, and also opening doors to investigate the strong-field and ultrafast carrier dynamics of this emerging material.
\end{abstract}

\section{Introduction}
Since the rediscovery of black phosphorus (BP) from the perspective of a two-dimensional (2D) layered material in 2014\cite{li_black_2014,xia_rediscovering_2014,liu_phosphorene_2014,churchill_two-dimensional_2014,koenig_electric_2014,rodin_strain-induced_2014,tran_layer-controlled_2014,qiao_high-mobility_2014}, significant amount of scientific and technological interest has been attracted to investigate this fascinating new material, with intensive research efforts ongoing and focusing on its numerous new properties and novel applications\cite{liu_semiconducting_2015,ling_renaissance_2015,Dhanabalan2017,yZhou2017,Xia2019}. Compared to other members of the 2D materials family, such as semimetallic graphene\cite{novoselov_two-dimensional_2005,zhang_experimental_2005}, semiconducting transitional metal dichalcogenides (TMDs)\cite{mak_atomically_2010,radisavljevic_single-layer_2011}, and insulating hexagonal boron nitride (hBN)\cite{Novoselov2005}, BP has many unique and attractive properties. Thin film BP exhibits a moderate and direct band gap within an appealing energy range\cite{li_black_2014}, and thus can bridge the energy gap between that of gapless graphene and relatively large-bandgap TMDs. Measurements of electric and optical properties of BP imply that it is a promising candidate for novel optoelectronic and photonic applications in the midinfrared to visible frequency range\cite{xia_rediscovering_2014,rzhang2015,xyzhou2015,xyzhou2015b,mao_optical_2016,lkLi2016,fyYang2018,deng_progress_2018}. Devices based on the linear optical response of BP, such as photodetectors\cite{youngblood_waveguide-integrated_2015,yuan_polarization-sensitive_2015,guo_black_2016,chen_widely_2017} and optical modulators\cite{whitney_field_2017}, have been reported. Besides, the nonlinear optical properties of BP in the perturbative regime have also been studied, demonstrating its great potential in light absorption\cite{lu_broadband_2015}, optical modulation\cite{zheng_few-layer_2017}, and all-optical signal processing\cite{zheng_black_2017}, etc. Apart from these, BP presents a higher free-carrier mobility (about 1000 cm$^2$V$^{-2}$s$^{-1}$)\cite{li_black_2014,xia_rediscovering_2014} compared to other TMDs semiconductors such as MoS$_2$ (around 200 cm$^2$V$^{-2}$s$^{-1}$)\cite{radisavljevic_single-layer_2011}. With a combination of relatively high on-off ratio\cite{li_black_2014}, BP can thus find attractive applications in building high-performance high-frequency thin film electronics technology\cite{wang_black_2014}. Another notable feature of BP is its remarkable in-plane anisotropy nature, including anisotropic electronic, optical and phonon responses\cite{xia_rediscovering_2014,qiao_high-mobility_2014}, making it of particular interest for designing conceptually new nano-device applications, e.g., novel plasmonic devices with tunable resonance frequency\cite{low_plasmons_2014} and efficient thermo-electric device via orthogonal electric and heat transport directions\cite{fei_enhanced_2014}. Within the past five years, we have witnessed enormous exciting research progress demonstrating BP as a promising 2D material for both fundamental studies and various applications. However, the important extreme nonlinear behavior and ultrafast dynamics of atomically thin film BP under strong-field excitation have yet to be revealed.

In this paper, we report investigations of strong-field nonlinear optical properties of monolayer BP using first-principles simulations based on the framework of real-time time-dependent density-functional theory (TDDFT)\cite{Runge1984,Leeuwen1998,Castro2004a}. Nonperturbative high harmonic generation (HHG) has been observed in BP up to the 13th order driven by a midinfrared femtosecond laser pulse at intensity of $2\times 10^{11}$ W/cm$^2$. Compared to other typical monolayer materials in the 2D materials family, i.e., graphene, MoS$_2$ and hBN, BP shows remarkably superior HHG properties with both higher cutoff energy and significantly enhanced intensity. In addition, BP can exhibit strong in-plane anisotropy in HHG properties. This study advances the scope of the current research activities of thin film BP. The previously unidentified superior strong-field optical properties of BP may intrigue the community with novel applications such as developing conceptually new extreme-ultraviolet and/or attosecond photonic and optoelectronic nanodevices. It also offers an opportunity for understanding the ultrafast carrier dynamics of this emerging material. 

\section{Results}
\subsection*{Crystal and band structures}
\begin{figure}[htbp]
\centering
\includegraphics[width=0.8\textwidth
]{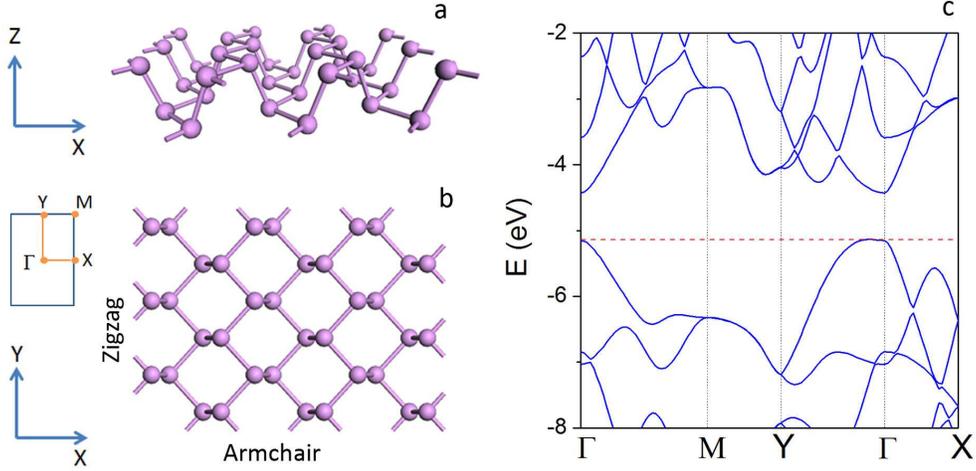}
\caption{\label{structure} Atomic structure and electronic band structure of monolayer black phosphorus (BP). (a) Side and (b) top view of the BP crystal lattice. $x$ and $y$ correspond to the armchair and zigzag directions, respectively. The associated Brillouin zone is also shown. (c) Calculated band structure of monolayer BP. A direct band gap around 0.72 eV is observed at the $\Gamma$-point.}
\end{figure}

Figure \ref{structure}(a) shows the side view of the crystal structure of a monolayer BP, which includes two atomic layers of phosphorus atoms, since each phosphorus atom is covalently bonded to three adjacent atoms to form a puckered geometry. The lattice top view is shown in Fig \ref{structure}(b), exhibiting a hexagonal honeycomb structure. The $x$ and $y$ directions, and the corresponding armchair (AC) and zigzag (ZZ) directions respectively, are denoted in the figure. The atomic structure of BP is optimized by using the CASTEP package\cite{castep2005} with the local density approximation (LDA) functional\cite{Marques2012}. The optimized in-plane lattice constants along the $x$ and $y$ directions are $a=4.50$ \AA~ and $b=3.24$ \AA, respectively. The inset shows the associated Brillouin zone (BZ), where the $\Gamma$-X and $\Gamma$-Y directions correspond to the AC and ZZ directions in real space, respectively.

Figure \ref{structure}(c) presents the electronic band structure of the monolayer BP. An impressive feature of highly asymmetric band dispersion around the $\Gamma$-point can be clearly seen, where both the valence and conduction bands are relatively flat along the $\Gamma$-Y direction but significantly dispersive along the $\Gamma$-X direction. Consequently, the effective mass of the free carriers is highly anisotropic in BP, which leads to strong in-plane anisotropy regarding its electronic, optical and phonon properties. The band energy is obtained by solving the Kohn-Sham equation using the Octopus package\cite{Andrade2015,Castro2006,Andrade2012} with the LDA functional, which results in a direct band gap of 0.72 eV at the $\Gamma$-point. This is lower than the experimental result due to the known underestimated-band-gap problem of the LDA functional. Using more advanced exchange-correlation functional could improve the band structure. However, this does not change our HHG results significantly, as we will show later. For example, TB09 meta generalized gradient approximation (MGGA) functional\cite{tb09} has been tested which gives a more accurate band gap of 1.5 eV. 

\subsection*{Nonperturbative HHG}
\begin{figure}[htbp]
\centering
\includegraphics[width=0.8\textwidth
]{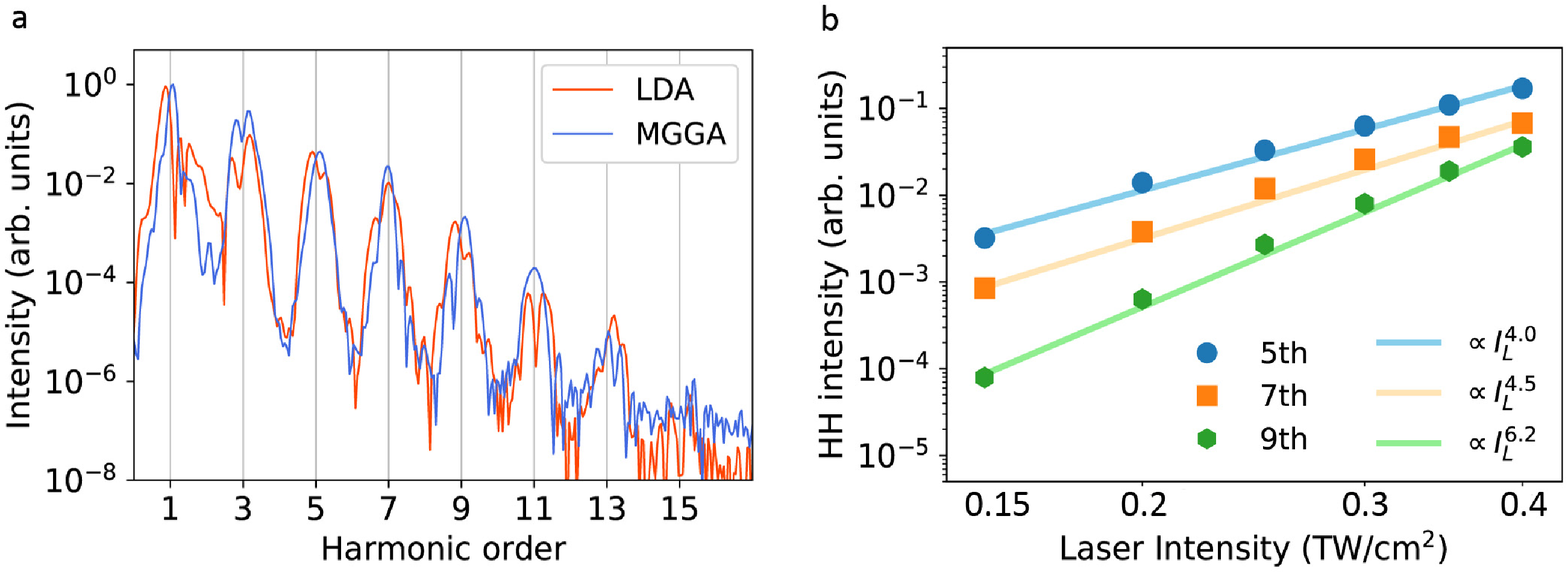}
\caption{\label{nonperturbative} Nonperturbative high harmonic (HH) generation in monolayer BP. (a) HH spectra driven by a laser with wavelength of 1.6 $\mu$m, intensity of $2\times 10^{11}$ W/cm$^2$, and polarization along the $x$ (armchair) direction. The compared spectra are obtained with different exchange-correlation functional, i.e., local density approximation (LDA) and TB09 meta generalized gradient approximation (MGGA). (b) Calculated HH intensity as a function of the peak pump intensity $I_L$ for representative harmonic orders $n=$5 (circles), 7 (squares), and 9 (hexagon). Solid lines are fitting of the calculated data to power laws, yielding exponents of 4.0, 4.5 and 6.2 for the 5th, 7th, and 9th harmonic, respectively, showing nonperturbative scaling for the HH process.}
\end{figure}

HHG in solids is an emerging research field\cite{Ghimire2011,Schubert2014,Vampa2015a,Hohenleutner2015,Ndabashimiye2016,Kruchinin2018} that provides exciting opportunities to study strong-field and ultrafast dynamics in the condensed phase\cite{Vampa2014,Vampa2015b,Luu2015,You2017,TD2017a,Luu2018} as well as a novel approach to develop compact extreme-ultraviolet and attosecond photonics\cite{TD2017b,Langer2017,Hammond2017,Sivis2017,Garg2018}. 
HHG in 2D materials is of particular interest as the 2D system exhibits distinctive electronic properties and the study may lead to a new generation of nanoelectronic and nanophotonic devices\cite{Liu2017,Cox2017,Yoshikawa2017,Taucer2017,Baudisch2018,TD2018}. 
Figure \ref{nonperturbative}(a) shows a representative high harmonic spectrum generated in monolayer BP at a peak pump intensity of $2\times 10^{11}$ W/cm$^2$ and wavelength of 1600 nm. Well-defined spectral peaks at integer multiples of the fundamental pump frequency up to the 13th order can be observed. Only odd harmonic orders are present, reflecting the centrosymmetric nature of the crystal lattice. The results of simulations with different exchange-correlation functional, i.e., LDA and TB09 MGGA, are compared in Fig. \ref{nonperturbative}(a), which show the HHG spectra are not affected significantly by the choice of exchange-correlation functional. The difference in peak spectral intensity is less than a factor of three, which will not impact our conclusions. Therefore, the LDA functional is used throughout the work unless otherwise specified. 

Figure \ref{nonperturbative}(b) shows the calculated harmonic intensity as a function of the pump laser intensity $I_L$. Power-law fit to the data yield intensity dependence of $I_{L}^{4.0}$, $I_{L}^{4.5}$, and $I_{L}^{6.2}$ for the 5th, 7th, and 9th harmonic, respectively, which clearly show the nonperturbative character of the HHG process in this study. In contrast, in the perturbative limit, the intensity scaling for the $n$th harmonic would be $I_{L}^{n}$.

We have also check that the HHG can operate at a wide range of laser wavelength, including the longer mid-infrared region\cite{jfLi2016}. The cutoff photon energy of the high harmonics for different driving wavelength is at the same level. Apart from monolayer BP, HHG in few-layer BP can also be expect, albeit may with reduced efficiency per layer\cite{Liu2017,le_breton_high-harmonic_2018}.

\subsection*{HHG mechanism}
\begin{figure}[htbp]
\centering
\includegraphics[width=0.75\textwidth
]{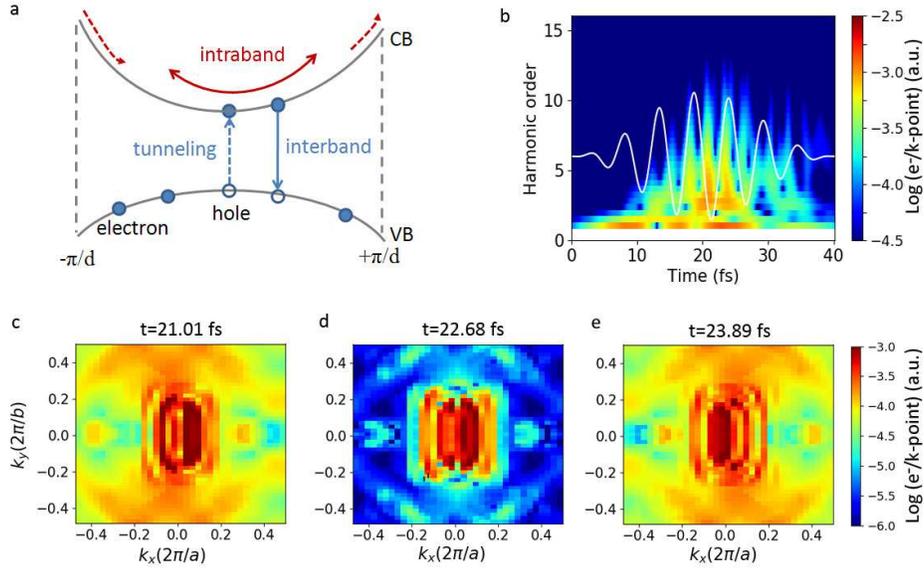}
\caption{\label{wavelet} (a) Schematically illustration of the two mechanisms, i.e., interband and intraband contribution, responsible for high harmonic generation (HHG) in solids. (b) Time-frequency analysis of the HHG process. The white curve is the profile of the applied vector potential. (c-e) Subcycle dynamics of the excited electrons in the momentum space at time t = 21.01 fs, 22.68 fs, and 23.89 fs, respectively. The in-phase signature of HHG with the driving fields in panel (b) and the oscillation features of excited electrons in the Brillouin zone in panels (c-e) show evidence of intraband process as the dominant HHG mechanism in this study.}
\end{figure}

Two distinctive mechanisms, i.e., the interband and intraband mechanisms, have been identified to be responsible for HHG in solids, as schematically illustrated in Fig. \ref{wavelet}(a). In the interband picture, electron-hole pairs are generated, and then recombined to emit HH photons after accelerated by the laser field. Thus the interband HHG is associated with recolliding process born before the peak field. On the other hand, in the intraband process, acceleration of charge carriers driven by the laser field in anharmonic bands give rise to harmonic emission. Thus the intraband harmonics are emitted in phase with the laser field. Therefore, time-frequency analysis of the HHG process may shed light on the underlying HHG mechanisms\cite{Vampa2017}. Figure \ref{wavelet}(b) shows the wavelet spectrogram, which reveals HHG in BP as discrete bursts in phase with laser peaks corresponding to maximum carrier acceleration. This in-phase signature instead of recombination trajectories shows strong evidence that supports intraband contribution as the dominate HHG mechanism in BP for the parameters we study here. 

As significant intraband contribution implies electrons traversing a great part of the BZ, we then analysis the BZ explored by the electrons to gain further insights. We study the momentum space resolved subcycle dynamics of the excited electrons. Three typical cases are shown in Fig. \ref{wavelet} (c-e) corresponding to the time of t = 21.01 fs, 22.68 fs, and 23.89 fs, respectively. In all the cases, areas near the $\Gamma$-point have the highest number of excited electrons due to the small energy gap near the $\Gamma$-point. When t = 22.68 fs, excited electrons are concentrated near the $\Gamma$-point, which corresponds to the time when the system has the lowest excitation. For t = 21.01 fs and 23.89 fs, a large part of the BZ has significant amount of excited electrons, and more importantly, the excited electrons evidently arrange from the left to the right boundaries of the BZ. These subcycle dynamics clearly show the oscillation of excited electrons driven by the laser field, which is consistent with the above time-frequency analysis of the intraband process in BP that contributes to the HHG. 

\subsection*{Superior HHG properties}
\begin{figure*}[htbp]
\centering
\includegraphics[width=0.98\textwidth
]{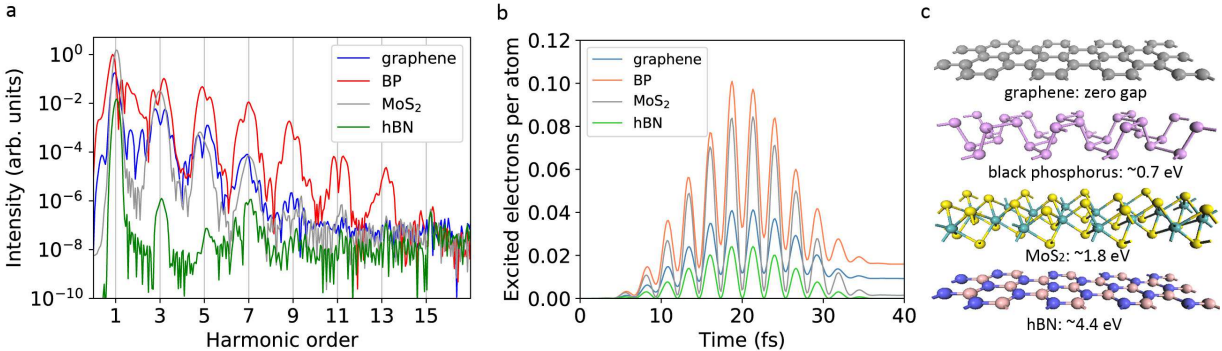}
\caption{\label{comparison} (a) Comparison of high harmonic generation in representative types of monolayer 2D materials, i.e., graphene (semimetal with zero gap), black phosphorus (semiconductor with computed 0.72 eV gap), MoS$_2$ (semiconductor with computed 1.82 eV gap), and hexagonal boron nitride (insulator with computed 4.4 eV gap). (b) Number of electrons excited to the conduction bands during the laser pulse. (c) Illustrative images of these representative 2D material samples. The laser parameters are the same as those in Fig. \ref{nonperturbative}(a).}
\end{figure*}

Current research of HHG in 2D materials has been primarily focused on graphene\cite{Cox2017,Yoshikawa2017,Taucer2017,Baudisch2018,Mikhailov2007,Mikhailov2008,Ishikawa2010,Avetissian2012a,Avetissian2012b,AN2014,AN2015,
Chizhova2016,Chizhova2017,Avetissian2018,cdLiu2018,chen_circularly_2019}, the most known prototype atomically thin material. Besides, studies of HHG in silicene\cite{Qin2018}, hBN\cite{TD2018,le_breton_high-harmonic_2018}, and MoS$_2$\cite{Liu2017,guan_cooperative_2019} have also been reported.
To see the superior HHG properties in BP, here we present comparison of HHG from representative types of monolayer 2D materials, i.e., graphene (semimetal with zero gap), BP (semiconductor with computed 0.72 eV gap), MoS$_2$ (semiconductor with computed 1.82 eV gap), and hBN (insulator with computed 4.4 eV gap), as shown in Fig. \ref{comparison}. The pump laser intensity is $2\times 10^{11}$ W/cm$^2$. The highest harmonic order observed in graphene is 9th, in agreement with the experimental result reported by Yoshikawa \textit{et al}\cite{Yoshikawa2017}. The HHG in MoS$_2$ is up to the 11th order, but the harmonic intensity is at the same level as that in graphene. Under the same pump condition, however, HHG signal in hBN is very weak. The harmonics are only up to the 7th order, and furthermore, the spectral intensity is lowered by several orders of magnitude. We note that in recent theoretical studies of HHG in monolayer hBN, HHG up to about 13th order are obtained driven by lasers with the same wavelength but with peak intensity 1-2 orders of magnitude higher\cite{TD2018,le_breton_high-harmonic_2018}. Thus it is reasonable to expect weak HHG response with our low pump intensity. In contrast, HHG in BP is remarkable, presenting superior properties compared to the others. The energy cutoff is the highest, extending to the 13th order. Moreover, the harmonic intensity is significantly higher than that in graphene, MoS$_2$, and hBN. The intensity enhancement reaches one order of magnitude for the 3rd and 7th harmonics, and two orders of magnitude for the rest of the presenting harmonic orders compared to that of graphene and MoS$_2$. The harmonic order and intensity may be further improved by optimizing laser parameters such as wavelength and intensities. The result here demonstrates the great potential of BP as an outstanding and promising candidate for developing novel optoelectronic and photonic nanodevices in the extreme-ultraviolet and/or attosecond regime. 

The superior strong-field nonlinear optical properties of BP is related to its unique band structure and ultrafast carrier dynamics. To illustrate the dynamics of excited carriers, we show the number of electrons excited to the conduction bands during the laser pulse for the three different samples in Fig. \ref{comparison}(b). The number of excited electrons in hBN is the lowest and thus has the lowest high harmonic intensity, as can be expected from an insulator with a large band gap. Besides, the number of excited electrons in graphene is also smaller than that of BP and MoS$_2$ despite graphene's zero band gap. This can be explained by the density of states (DOS) of the materials. Graphene has a vanishing DOS at the Fermi level, while BP and MoS$_2$ have a flat band dispersion near the Fermi level that can lead to a large DOS. Therefore more electrons from valence bands can be excited in BP and MoS$_2$. These carrier dynamics are in accordance with the HHG results shown in Fig. \ref{comparison}(a). Besides number of carriers, the intraband contribution also largely depends on the shape of band dispersion. BP and graphene exhibit (near) linear band dispersion near the Fermi level. In comparison, the band is relatively flat for MoS$_2$. Therefore, BP and graphene have a larger anharmonicity than MoS$_2$. Taken these into account,  MoS$_2$ has a HHG efficiency lower than BP but comparable to graphene.

\subsection*{Anisotropic HHG}
\begin{figure}[htbp]
\centering
\includegraphics[width=0.75\textwidth
]{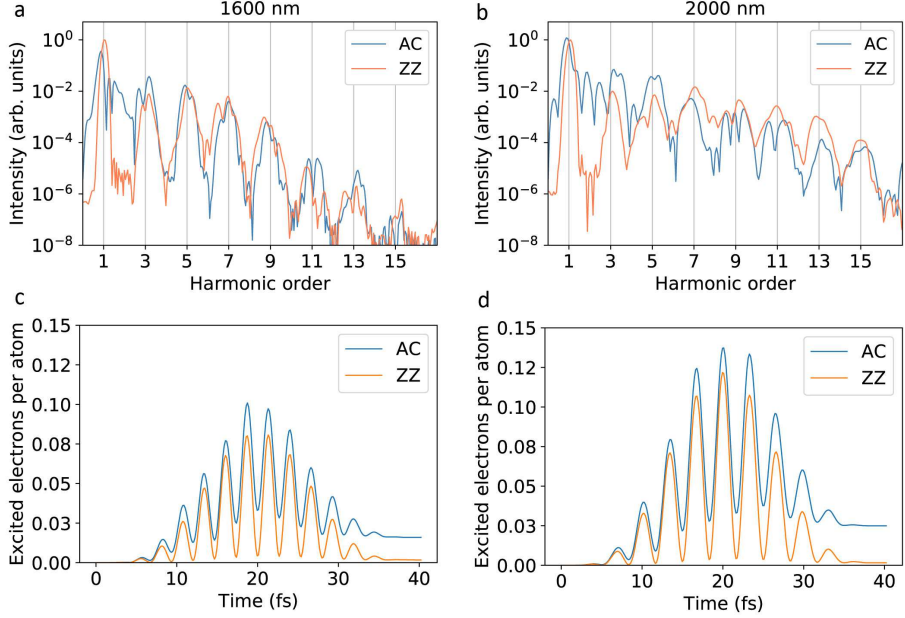}
\caption{\label{aniso} (a-b) High harmonic spectra driven by laser polarized along either the armchair (AC) or zigzag (ZZ) direction for two different wavelength of (a) 1600 nm and (b) 2000 nm. (c-d) Number of electrons excited to the conduction bands during the laser pulse corresponding to (a-b), respectively.}
\end{figure}

Another attractive feature of BP, distinguishing it from many other isotropic 2D materials such as graphene and MoS$_2$, is the remarkable in-plane anisotropic properties that may enable novel device applications. While many have studied the anisotropic properties in BP such as the electrical conductance\cite{xia_rediscovering_2014,qiao_high-mobility_2014}, thermal conductance\cite{fei_enhanced_2014}, and linear/nonlinear optical responses\cite{mao_optical_2016,youngblood2017}, etc., the anisotropic extreme nonlinear optical properties under strong-field excitations remain unexplored. Figure \ref{aniso} (a-b) show the HHG spectra with excitation laser polarized either along AC ($\Gamma $-X) or ZZ ($\Gamma $-Y) direction for two different laser wavelength, i.e., 1600 nm and 2000 nm, respectively. Evident anisotropy in harmonic intensity and spectral width for each harmonic can be observed. The harmonic yield excited by laser polarized along the AC direction is higher than that along the ZZ direction. The anisotropic characteristic between the two directions driven by the 2000 nm laser is more evident than that by the 1600 nm laser. The largest difference in peak harmonic intensity can reach one order of magnitude. To understand these results, we plot the computed number of electrons excited to the conduction bands during the laser pulse in Fig. \ref{aniso}(c-d), corresponding to Fig. \ref{aniso}(a-b), respectively. For both laser wavelength, we see that the number of excited electrons is larger for the AC-polarized laser, which can thus lead to stronger electronic current and higher harmonic yield along the AC direction. The difference in carrier excitation along different directions may be attributed to the strong anisotropic band dispersion (see Fig. \ref{structure}(c)) and electronic properties of BP. Note that the effective mass of carriers along the AC direction is about 10 times smaller than that along the ZZ direction. As for the 2000 nm laser exciting more electrons to the conduction bands for both directions, it may be attributed to a larger ponderomotive potential that electrons feel in a longer-wavelength laser field. 

\section{Conclusions}
In summary, we have explored monolayer BP in the new regime of strong-field nonlinear optical responses. It opens a door to investigate strong-field nonlinear behavior and ultrafast carrier dynamics of this emerging material system. Nonperturbative HHG in BP exhibits remarkable properties with extended cutoff energy and orders of magnitude higher intensity than that of graphene, MoS$_2$, and hBN. Besides, the anisotropy of HHG in BP has been observed and the carrier dynamics analysed. These findings provide additional merits to monolayer BP that can lead to promising new applications, e.g., enabling significant performance improvements of next generation optoelectronic and photonic nanodevices in the extreme-ultraviolet and attosecond regime. The time scale of laser-induced ultrafast electron dynamics occurs in the attosecond regime. The spatial scale of light can be confined into nanoscopic volumes in nanophotonics. The combination of attosecond and nanophotonics techniques will allow the manipulation and control of light and associated electron dynamics with attosecond-nanometre resolution, in addition to the development of compact attosecond sources at nanoscale. We foresee that some novel devices based on HHG in BP may emerge in the future, such as compact and efficient frequency converters, petahertz electron signal processors, coherent nano-imaging with attosecond temporal resolution, and UV/XUV nanoprobes in nano-medicine. It is also worthy of noting that naked BP in air is less stable than other widely used 2D materials such as graphene and TMDs. Material stability could be one major concern of the BP-based devices. However, the BP material can show good stability if isolated from water and oxygen. To date, many effective passivation and packaging methods have been demonstrated that may overcome the degradation issue, e.g., BP surface passivation via metal-ion modification is shown to enhance both the stability and transistor performance of BP sheets\cite{zGuo2017}; BP encapsulated between hBN layers also shows long-term stability in ambient condition even under strong laser irradiation\cite{yCao2015}.  

\section{Methods}
The BP structure is studied by a semiperiodic supercell model, which is a rectangular cell containing four phosphorus atoms. A vacuum space of 30 Bohr, which includes 3 Bohr of absorbing regions on each side of the black phosphorus monolayer, is chosen to eliminate the interactions between adjacent layers and avoid unphysical reflection error in the spectral region of interest. We optimize the black phosphorus structure by using the CASTEP package\cite{castep2005}, and the Octopus package\cite{Andrade2015,Castro2006,Andrade2012} is used for other calculations throughout this work. The ground state geometric structure and electronic properties was determined within the density functional theory (DFT) framework in the local density approximation (LDA)\cite{Marques2012}. 

Time evolution of the wave functions and time-dependent electronic current are studied by propagating the Kohn-Sham equations in real time and real space\cite{Castro2004b} within the time-dependent DFT (TDDFT) framework\cite{Runge1984,Leeuwen1998,Castro2004a} in the adiabatic LDA (ALDA). The real-space spacing is 0.46 Bohr. A $36\times 48 \times 1$ Monkhorst-Pack k-point mesh for Brillouin zone sampling is used. The fully relativistic Hartwigsen, Goedecker, and Hutter (HGH) pseudopotentials are used in all our calculations. 

The laser is described in the velocity gauge, which has a wavelength of $\lambda_L=$ 1600 nm (corresponding to a photon energy of 0.77 eV) and FWHM pulse duration of $\tau$ = 20 fs. The peak laser intensity in vacuum is $I_L=2\times10^{11}$ W/cm$^{2}$. The pulse has a sin-squared envelop profile and the carrier-envelope phase is taken to be $\Phi= 0$. The monolayer BP is excited at normal incidence so that the driving electric field is in the $x-y$ plane of the sample.

The HHG spectrum was calculated from the time-dependent electronic current $\textbf{j}(\textbf{r},t)$ as:
\begin{equation}
\mathrm{HHG}(\omega) = \Big| \mathcal{FT} \Big(\frac{\partial}{\partial t} \int \textbf{j}(\textbf{r},t) \ \mathrm{d}^3 \textbf{r}  \Big) \Big|^2,
\end{equation}
where $\mathcal{FT}$ denotes the Fourier transform. 

The total number of excited electrons was calculated by projecting the time-dependent Kohn-Sham states onto the the ground-state Kohn-Sham states. As the \textit{n}-th state evolves in time, it has some possibility to transit to other states and thus contain other ground-state components. The total number of excited electrons $N_{ex}(t)$ is calculated as:
\begin{equation}
N_{ex}(t)=N_e-\int_{\mathrm{BZ}} \sum_{n,m}^{\mathrm{occ}} f_{n,k} |\langle\phi_{n,k}(t)|\phi_{m,k}(0)\rangle|^2 \mathrm{d} \textbf{k}
\end{equation}

where $N_e$, $\phi_{n,k}(t)$,$\phi_{m,k}(0)$, $f_{n,k}$, BZ denote the total number of electrons in the system, the time-dependent Kohn-Sham state at \textit{n}-th band at k-point \textit{k}, the ground-state (t = 0) Kohn-Sham state at \textit{m}-th band at k-point \textit{k}, and the occupation of Kohn-Sham state at \textit{n}-th band at k-point \textit{k}, integration over the whole Brillouin zone, respectively.

The momentum-resolved number of excited electron $N_{ex}(k,t)$ is calculated as:
\begin{equation}
N_{ex}(k,t)=\sum_{m}^{\mathrm{unocc}}\sum_{n}^{\mathrm{occ}} f_{n,k} |\langle\phi_{n,k}(t)|\phi_{m,k}(0)\rangle|^2
\end{equation}



\bibliography{bp-ref}

\begin{acknowledgement}
We acknowledge financial support from the National Natural Science Foundation of China (NSFC) (11705185) and the Presidential Fund of China Academy of Engineering Physics (CAEP) (YZJJLX2017002). 
\end{acknowledgement}



\end{document}